\documentclass[pdftex,12pt]{article}

\usepackage{scicite}
\usepackage{graphicx}
\usepackage{times}
\usepackage{color}

\newenvironment{sciabstract}{%
\begin{quote} \bf}
{\end{quote}}

\title{Phonon hydrodynamics and ultrahigh-room-temperature thermal conductivity in thin graphite}

\author
{Yo Machida,$^{1\ast}$ Nayuta Matsumoto,$^{1}$ Takayuki Isono,$^{1}$ and Kamran Behnia$^{2\ast}$\\
\\
\normalsize{$^{1}$Department of Physics, Gakushuin University, Tokyo 171-8588, Japan}\\
\normalsize{$^{2}$Laboratoire Physique et Etude de Mat\'{e}riaux (CNRS-Sorbonne Universit\'e-ESPCI),}\\
\normalsize{PSL Research University, 75005 Paris, France}\\
\\
\normalsize{$^\ast$To whom correspondence should be addressed;}\\
\normalsize{ E-mail: yo.machida@gakushuin.ac.jp; kamran.behnia@espci.fr.}
}

\date{January 17, 2020}

\begin{document} 


\baselineskip24pt


\maketitle 



\begin{sciabstract}

Allotropes of carbon, such as diamond and graphene, are among the best conductors of heat. We monitored the evolution of thermal conductivity in thin graphite as a function of temperature and thickness and found an intimate link between high conductivity, thickness, and phonon hydrodynamics. The room temperature in-plane thermal conductivity of 8.5-micrometer-thick graphite was 4300 watts per meter-kelvin-a value well above that for diamond and slightly larger than in isotopically purified graphene. Warming enhances thermal diffusivity across a wide temperature range, supporting partially hydrodynamic phonon flow. The enhancement of thermal conductivity that we observed with decreasing thickness points to a correlation between the out-of-plane momentum of phonons and the fraction of momentum relaxing collisions. We argue that this is due to the extreme phonon dispersion anisotropy in graphite.

\end{sciabstract}

Heat travels in insulators because of the propagation of collective vibrational states of the crystal lattice called phonons. The standard description of this transport phenomena invokes quasi-particles losing their momentum to the underlying lattice  due to collisions along their trajectory~\cite{ziman}. Gurzhi proposed decades ago that phonons in insulators and electrons in metals can flow hydrodynamically if momentum-conserving collisions among carriers become abundant~\cite{gurzhi}. Recently, Hydrodynamic regimes for electrons~\cite{moll,crossno,bandurin} and for phonons~\cite{lee,cepellotti,machida,martelli,huberman} has become a subject of renewed attention, partially driven by the aim of quantifying the quasi-particle viscosity.

Unlike particles in an ideal gas of molecules, the phonon momentum is not conserved in all collisions. When scattering between two phonons produces a  wave-vector exceeding the unit vector of the reciprocal lattice, the excess of momentum is lost to the underlying lattice. These are called Umklapp (U) scattering events and they require sufficiently large wave-vectors. Because cooling reduces the typical wavelength of thermally-excited phonons, U scattering rarefy with decreasing temperature and most collisions among phonons conserve momentum, becoming Normal (N) scattering events. In this context, a regime of phonon hydrodynamics emerges that is sandwiched between diffusive and ballistic regimes~\cite{gurzhi}. Observations of the hydrodynamic regime include several solids~\cite{machida,martelli,deglin,thomlinson,kopylov,zholonko}. In this narrow temperature window, warming multiplies normal collisions and this enhances the ratio of thermal conductivity to specific heat called the thermal diffusivity. Observations of this behavior tend to be at cryogenic temperatures.   

 The domination of N events over U events across very broad temperature range in graphene led two groups to propose that phonon hydrodynamics might be observed at temperatures outside the cryogenic range~\cite{lee,cepellotti}. However, heat transport measurements in graphene~\cite{nika} are challenging to study by the standard four-probe steady-state technique. Evidence for second sound, a manifestation of phonon hydrodynamics was recently found at temperatures exceeding 100 K in graphite~\cite{huberman}. These observations were in agreement with theoretical expectations~\cite{ding}.


The two-dimensional lattice of graphite (Fig.~1A, inset) consists of strong interlayer  sp$^2$ covalent bonds combined with weak intralayer van der Waals bonds. The strength of the in-plane and the out-of-plane couplings differ by  two orders of magnitude. This dichotomy makes graphite easily cleavable down to the single-layer graphene form~\cite{geim}. The bonding of graphite also creates two distinct Debye temperatures, one for the in-plane and the other for the out-of-plane atomic vibrations~\cite{krumhansl}. This induces a large difference between in-plane and out-of-plane thermal conductivities~\cite{slack}. The experimentally measured thermal conductivity  ~\cite{slack,bowman,holland,taylor,morelli} shows a roughly similar temperature dependence. However, there is a large variety in the reported magnitude of in-plane thermal conductivity, which at room temperature can vary between 72 and 2100 W/Km~\cite{slack}, a feature attributed to the unavoidable presence of the stacking faults and contamination of the in-plane data by a contribution from $c$-axis flow. As we will see below, new insight is provided by a thickness-dependent study on the same sample.

We measured the in-plane thermal conductivity ($\kappa$) of a commercially available highly oriented pyrolytic graphite (HOPG) samples, all peeled from a thick mother, with a standard steady-state one-heater-two-thermometers technique in high vacuum (Fig.~1). We tested the reliability by measuring the thermal conductivity of a long thin silver foil with a thermal resistance comparable to our most thermally resistive sample  and quantifying the small deviation from the Wiedemann-Franz law~\cite{SM}. For samples with thicknesses ranging from 8.5 $\mu$m to 580 $\mu$m, we found identical $\kappa$ behavior below 20 K and a steady thickness evolution for $\kappa$ with increasing temperature above 20 K. 

We compared the temperature dependence of $\kappa$ in the thickest sample (580 $\mu$m) with the measured specific heat (Fig.~2A). We found that $\kappa$ peaks around 100 K, similar to other measurements~\cite{bowman,holland,morelli}. Below this maximum, $\kappa$ quickly decreases and roughly follows a $T^{2.5}$ dependence, close to the specific heat trend below 10 K~\cite{alexander}. The specific heat ($C$) temperature behavior deviated from the $T^3$ expected from the Debye approximation. However, this behavior is not strictly observed in most real solids due to unequal distribution of phonon weights. The 2.5 exponent has been attributed in graphite to an admixture of $T^3$ and $T^2$ contributions by out-of-plane and in-plane phonons, respectively~\cite{komatsu}. This unusual exponent may have obscured the Poiseuille regime, which is usually associated with faster than cubic thermal conductivity~\cite{gurzhi}.  

Closer examination of the parallel evolution of thermal conductivity and specific heat can help unveil the Poiseuille regime as $\kappa$ evolves faster than $C$ above 10 K and slower below 10 K (Fig.~2A). Plotting $\kappa/T^{2.5}$  and $C/T^{2.5}$ makes this difference easier to recognize (Fig.~2B). Upon warming, $\kappa/T^{2.5}$ shows a pronounced maximum above 10 K, while $C/T^{2.5}$ gradually decreases. 
The thermal diffusivity, $D_{\rm th}$, is the ratio of thermal conductivity to specific heat (expressed in proper units of J/Kmol). We found $D_{\rm th}$ has  a non-monotonic temperature dependence and between 10 K and 20 K (Fig.~2C). The phonon hydrodynamic picture provides a straightforward interpretation of this feature. Warming leads to enhanced momentum exchange among phonons, because the fraction of collisions which conserve momentum increases. As a consequence, heat diffusivity rises. If all phonons had the same mean-free-path irrespective of their branch and wave-vector, this would also imply a rise in the effective mean-free-path. The Poiseuille maximum around 40 K and the Knudsen minimum around 10 K where diffuse boundary scattering rate is effectively increased because of N scattering, define the boundaries of this hydrodynamic window. 

We found that the electron contribution is negligibly small in the temperature range of interest by determining the Lorenz ratio ($L/L_0$). We measured electrical conductivity, $\sigma$, in order to quantify  $L=\frac{\kappa}{\sigma T}$ and compare it with $L_0=2.44\times10^{-8}$ W$\Omega$/K$^2$. This results in a ratio between 100 and 1000 above the Knudsen minimum (Fig.~2D).

The behavior we observed for $\kappa$ and $C$ are not due to outstanding sample quality. Comparable features can be found in published data~\cite{bowman,holland,SM}, but appear to have gone unnoticed. Our mother sample was an average HOPG containing both stable isotopes of carbon, ($\sim99\%$ $^{12}$C, $\sim1\%$ $^{13}$C). Our results support the conjecture that phonon hydrodynamics can occur without isotopic purity~\cite{machida}.

We measured an increased $\kappa$ as we decreased sample thickness (Fig.~3A). We performed successive  measurements after changing the thickness ($t$) of the sample along the $c$-axis, maintaining the sample width ($w=350$ $\mu$m)  and the distance ($l$) between contacts for the thermal gradient to be long enough compared to the thickness ($l/t>10$)~\cite{SM}. The trend is the opposite of observations for black phosphorus~\cite{machida}. With respect to the hydrodynamic regime, thinning leads to an amplification of the non-monotonic behavior of thermal diffusivity. This drives the Poiseuille peak to become sharper and towards higher temperatures.  Eventually, $D_{\rm th}$ of the thinnest sample shows a sharp maximum at 100 K. Second sound in graphite was observed near this temperature~\cite{huberman}. The thickness dependence vanishes below 10 K, presumably because the phonon mean-free-path in this range is set by the average crystallite size~\cite{slack}, which does not depend on thickness. Another possible origin of the thickness-independent low temperature thermal conductivity is an intrinsic  scattering of phonons by mobile electrons.

The thermal conductivity in our 240 $\mu$m thick sample is in reasonable agreement with previous observations on a similar thickness graphite~\cite{taylor}. The in-plane $\kappa$ we measured for the 8.5 $\mu$m  sample was $\sim$ 4300 W/Km. This exceeds the value for an isotopically pure graphene sample~\cite{chen} and higher than other bulk solids. The value is twice the value of natural abundance diamond ~\cite{wei} and about three times larger than high-purity crystalline BAs~\cite{kang,li,tian}. At room temperature, reducing thickness by two orders of-magnitude leads to a five-fold increase in $\kappa$ (Fig.~3C). Although the $\kappa$ we measured  is already  comparable with the highest values reported  in single-layer graphene ($\kappa\approx$ 3000 to 5000 W/Km)~\cite{chen,balandin}, our data does not saturate in the low thickness limit. In contrast to suspended graphene over a trench of 3 $\mu$m~\cite{balandin}, our samples are millimetric in length. Given the quasi-ballistic trajectory of phonons, we make the reasonable assumption that in-plane dimensions matter in setting the amplitude of thermal conductivity. This would imply that the ceiling is higher than previously believed and thinner samples with larger aspect ratio should display even larger conductivity. While several theoretical works have  predicted a robust hydrodynamic regime in graphene~\cite{lee,cepellotti} and its persistence in graphite~\cite{ding}, none examined the issue of thickness dependence.

To try to understand the origin of our observation, we scrutinized the occurrence of U and N collisions given the phonon dispersion of graphite~\cite{nihira,nika} (Fig.~4). We show the calculation of Nihira and Iwata~\cite{nihira} from a semicontinuum model for the in-plane and out-of-plane dispersion of longitudinal (LA), in-plane transverse (TA) and out-of-plane transverse (ZA) acoustic phonons along the $\Gamma$M and $\Gamma$A directions (Fig.~4B). The model parameters (velocities and elastic constants) were determined by using the best account of experimental specific heat data from 0.5 K to 500 K~\cite{nihira}. The two orientations show a striking contrast regarding the typical wavelength of thermally-excited phonons and requirements for U scattering. At 300~K (or 200 cm$^{-1}$), the typical in-plane wave-vector of the LA mode is only 0.1 of the BZ width. This makes U collisions extremely rare (Fig.~4C), because in order to create a phonon with a wave-vector larger than half of the BZ width, the average wave-vector of each colliding phonon needs to be 0.25 of the BZ width. The fundamental reason behind the scarcity of U collisions and the emergence of hydrodynamics resides this simple feature. The situation is radically different for out-of-plane wave-vectors. Even at 50~cm$^{-1}$, a thermally-excited phonon can have an out-of-plane wave-vector which is one-fourth of the BZ height. Above 90~cm$^{-1}$ (corresponding to 130~K), out-of-plane phonons are all thermally excited~\cite{nihira} and their peak wavelength is half of the BZ height. Any additional momentum along this orientation can kick them out of the BZ. A small $c$-axis component in the momentum exchanged by colliding phonons suffices for the collision to become a U event (Fig.~4D) and the heat flow to degrade.

Our observation implies a reduction in the relative weight of U collisions as the sample is thinned, since attenuating the relative rate of U collisions would extend the hydrodynamic window and enhance thermal conductivity. We note that the spacing between discrete available states in the reciprocal space depends on thickness. Therefore, the total number of states with out-of-plane momentum is inversely proportional to the thickness. It is true that only a small fraction of the Brillouin zone is wiped out by the finite size. However different collision mechanisms are competing for phase space and reducing the thickness not only reduces the population of the out-of-plane phonons, but also amplifies boundary scattering. Heat-carrying phonons can suffer either a U collision with an out-of-plane phonon, or a (more or less) specular collision at the boundary. Thus, reducing the thickness, by substituting a fraction of U collisions with specular boundary reflection, would limit the degradation of the heat flow. 

A satisfactory account of thickness dependence of thermal conductivity in both HOPG and  black phosphorus~\cite{machida} is missing.  Scattering at the boundaries and  imperfect transmission  across interfaces between  partially twisted graphene layers are to be put under scrutiny. Serious theoretical calculations are needed  to explain our findings.

\noindent
\textbf{Acknowledgments}\\
We thank S. Kurose for the contribution in the early stage of this study and A. Subedi for discussions. We also thank M. Tsubota and M. Watanabe for the technical supports. 
\textbf{Funding:} This work was supported by the Japan Society for the Promotion of Science Grant-in-Aids KAKENHI 16K05435, 17KK0088, and 19H01840 and also by the Agence Nationale de la Recherche (ANR-18-CE92-0020-01). 
\textbf{Author contributions:} Y.M. and K.B. conceived of and designed the study. Y.M., N.M., and T.I. performed the transport and specific heat measurements. Y.M. and K.B. wrote the manuscript with assistance from all the authors.
\textbf{Competing interests:} The authors declare no competing interests.
\textbf{Data and materials
availability:} All data is available in the manuscript or the supplementary material.
  
\section*{Supplementary materials}
Materials and Methods\\
Supplementary Text\\
Figs. S1 to S8\\
Table S1\\
References

\clearpage

\begin{figure}[p]
\vspace*{-10cm}
\includegraphics[width=16cm]{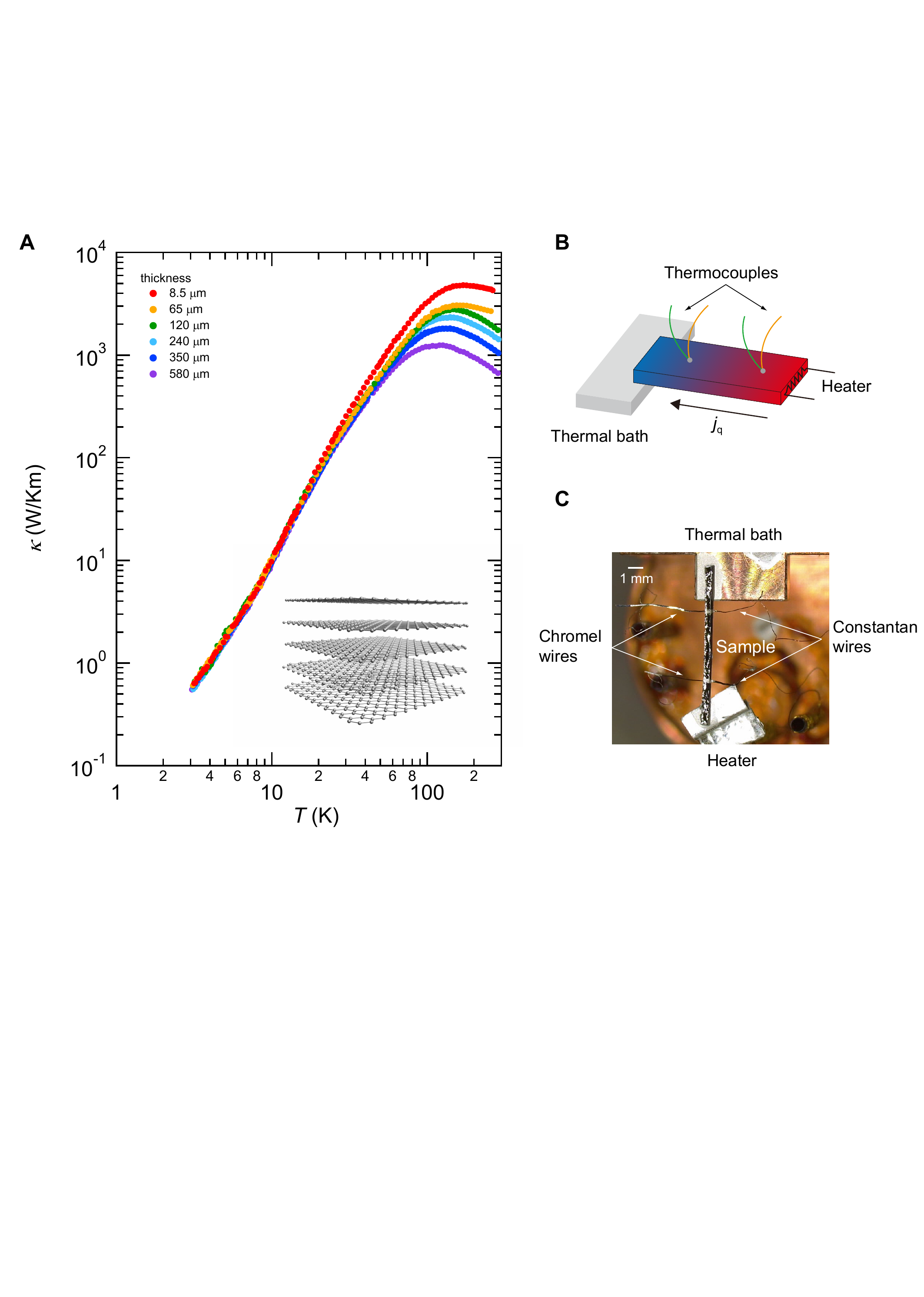}
\vspace*{-8cm}\\
{\bf Fig. 1. Thermal conductivity and experimental setup.} 
({\bf A}) Temperature dependence of in-plane thermal conductivity of graphite with various thickness ranging from 580 to 8.5 $\mu$m in a logarithmic scale. Inset shows side view of the crystal structure of graphite. 
A schematic illustration ({\bf B}) and a photo ({\bf C}) of the measurement setup for the thermal conductivity. Heat current $j_q$ generated by heater on the one end of the sample pass through the sample towards thermal bath. Temperature difference developed in the sample is determined by two pairs of thermocouples.
\end{figure}

\clearpage

\begin{figure}[p]
\vspace*{-5cm}
\begin{center}
\includegraphics[width=18cm]{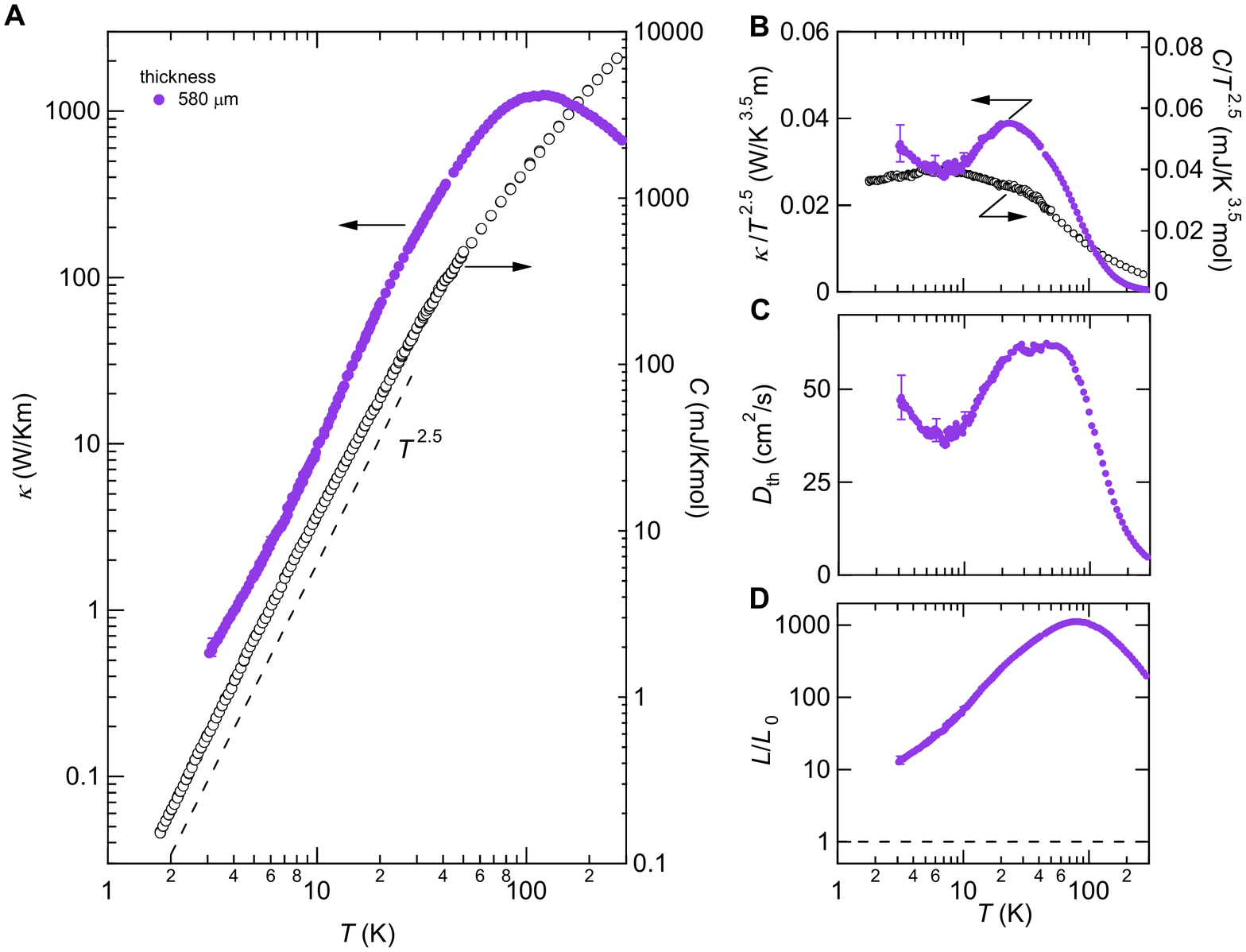}
\end{center}
\vspace*{0cm}
{\bf Fig. 2. Hydrodynamic heat transport.} 
({\bf A}) Temperature dependence of in-plane thermal conductivity $\kappa$ (left axis) and specific heat $C$ (right axis) of the 580 $\mu$m thickness graphite sample. 
({\bf B}) $\kappa$ divided by $T^{2.5}$ (left axis) and $C$ divided by $T^{2.5}$ (right axis) as a function of temperature. A pronounced maximum is seen only in $\kappa/T^{2.5}$ above 10 K. This yields a maximum in temperature dependence of thermal diffusivity $D_{\rm th}$ ({\bf C}).
Dominant phonon contribution in $\kappa$ is indicated by a large Lorenz ratio $L/L_0$ shown in ({\bf D}).
\end{figure}

\clearpage

\begin{figure}[p]
\begin{center}
\includegraphics[width=17cm]{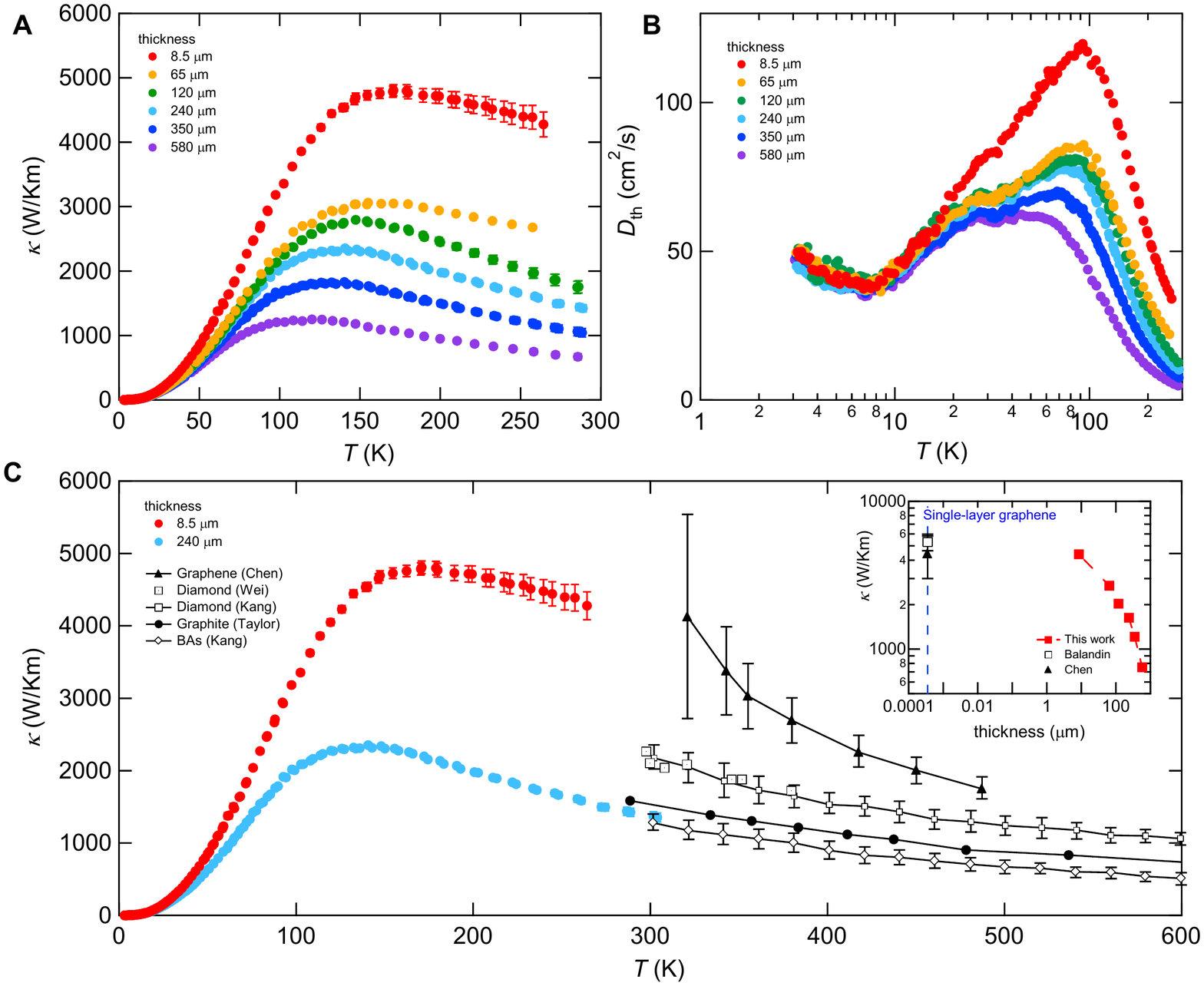}
\end{center}
\vspace*{-0cm}
{\bf Fig. 3. Thickness dependence of thermal conductivity.} 
({\bf A}) Temperature dependence of in-plane thermal conductivity $\kappa$ for various thickness of the samples. 
In the thinnest sample, $\kappa$ attains the largest value $\sim$ 4300 W/Km known in any bulk systems near the room temperature. 
({\bf B}) Temperature dependence of thermal diffusivity $D_{\rm th}$ for the various sample thickness. The maximum in $D_{\rm th}$ grows into a sharp
single peak with decreasing the thickness.
({\bf C}) Our data are compared with those of ultrahigh-thermal conductivity materials~\cite{chen,wei,taylor,kang}. The inset shows thickness dependence of thermal conductivity at 250 K. $\kappa$ of the thinnest sample is comparable with the high values reported in single-layer graphene~\cite{chen,balandin}.
\end{figure}

\clearpage

\begin{figure}[p]
\vspace*{-0cm}
\begin{center}
\includegraphics[width=16cm]{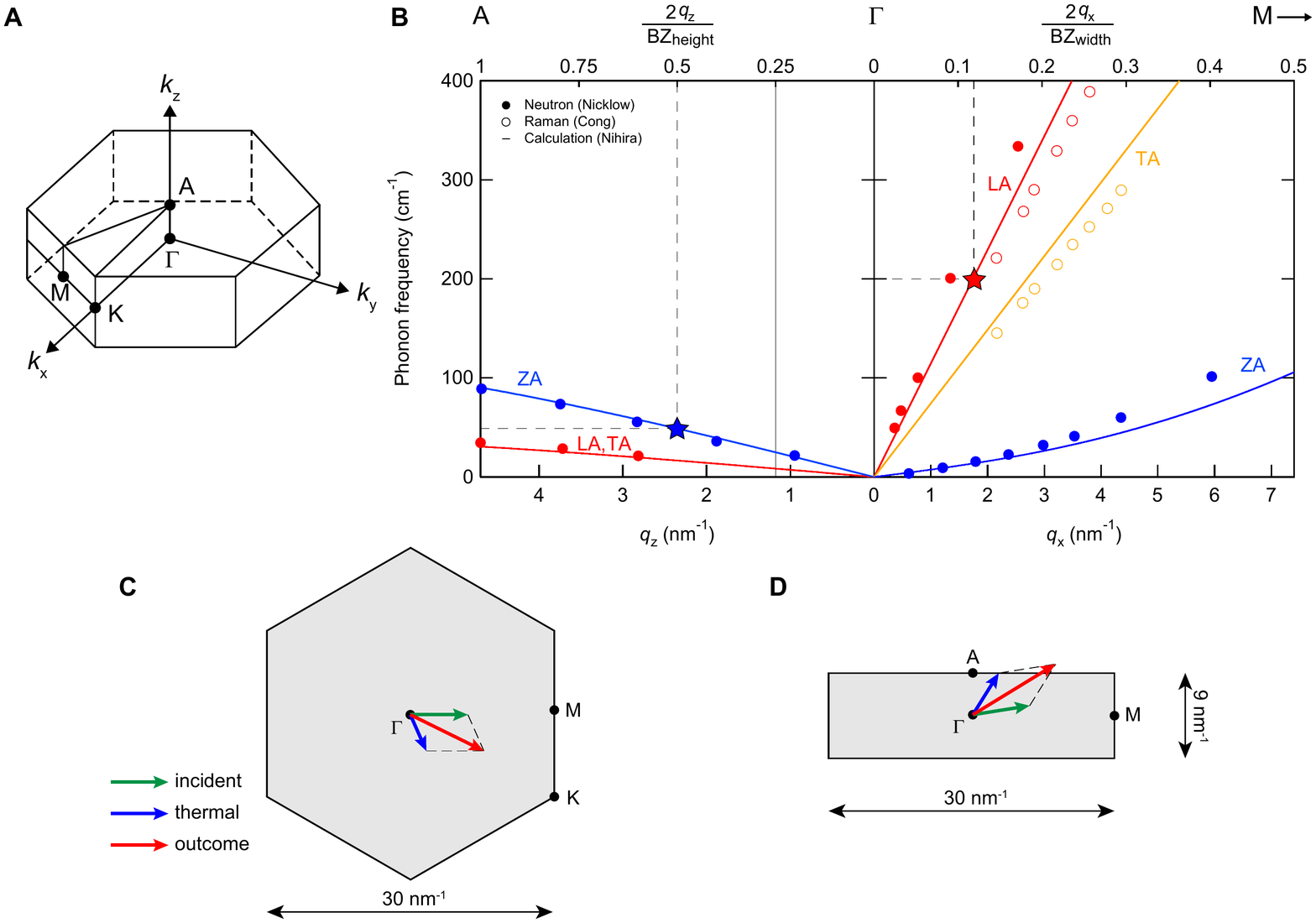}
\end{center}
\vspace*{0cm}
{\bf Fig. 4. Phonon dispersions.} 
({\bf A}) First Brillouin zone (BZ) of graphite.
({\bf B}) Calculated dispersions of acoustic phonon blanches along the $\Gamma$A and $\Gamma$M directions of BZ~\cite{nihira}, together with the experimental data obtained by neutron~\cite{nicklow} and Raman scattering~\cite{cong}.
BZ in the $\Gamma$KM plane ({\bf C}) and $\Gamma$MA plane ({\bf D}). Collision between the in-plane component of an incident phonon (green arrow) and a thermally excited phonon (blue arrow) remains N, because the in-plane wave-vector of thermal phonon is only a small fraction of the BZ width even at 300 K (or 200 cm$^{-1}$). Hence, the  wave-vector of outcome phonon (red arrow) does not exceed half of the BZ width. On the other hand, the out-of-plane wave-vector of a thermal phonon is one-fourth of the BZ height for frequencies as low as 50 cm$^{-1}$. Therefore, the collision becomes U, if the in-plane traveling phonon happens to possess a small out-of-plane component.

\end{figure}
\clearpage
\begin{center}
{\large Supplementary Materials for\\[0.3cm]
\bfseries{Phonon hydrodynamics and ultrahigh-room-temperature thermal conductivity in thin graphite}}

\vspace{0.5cm}

Yo Machida,$^{1\ast}$ Nayuta Matsumoto$^{1}$, Takayuki Isono$^{1}$, and Kamran Behnia$^{2\ast}$\\

\vspace{0.4cm}

\normalsize{$^{1}$Department of Physics, Gakushuin University, Tokyo 171-8588, Japan}\\
\normalsize{$^{2}$Laboratoire Physique et Etude de Mat\'{e}riaux (CNRS-UPMC), ESPCI Paris,}\\
\normalsize{PSL Research University, 75005 Paris, France}\\
\vspace{0.5cm}
\normalsize{$^\ast$To whom correspondence should be addressed;}\\
\normalsize{E-mail: yo.machida@gakushuin.ac.jp; kamran.behnia@espci.fr.}

\end{center}

\vspace{0.5cm}

\noindent
Materials and Methods\\
Supplementary Text\\
Figs.~S1 to S8\\
Table~S1\\
References

\clearpage

\section*{Materials and Methods}
\subsection*{Samples}
Sample is highly oriented pyrolytic graphite (HOPG) of the commercially available grade provided by TipsNano in Estonia.
A mother crystal with a dimension of 10 $\times$ 10 $\times$ 2 mm$^3$ is cut into a bar shape with an area of
10 $\times$ 0.35 mm$^3$ within the plane and thickness of 2 mm along the out-of-plane direction. Sample thickness was reduced by cleaving
while keeping the sample width (0.35 mm) unchanged and a ratio of distance between contacts to thickness large enough. In Table S1, sample
label represents a sequence of the experiments from the thickest to the thinnest sample.

\begin{table}[h]
{\bf Table S1. Sample dimensions.} Dimensions of the graphite samples (in mm), a ratio of distance between contacts to thickness ($l/t$),and a surface area $S$.\\
  \begin{center}
    \begin{tabular}{l|cccccc} \hline
      Sample label &1&2&3&4&5&6\\ \hline \hline
      Length (mm) & 10.0 & 10.0 & 8.2 & 8.2 & 5.1 & 5.9 \\
      Distance between contacts $l$ (mm) & 6.5 & 6.5 & 4.7 & 4.6 & 2.4 & 2.8 \\
      Width $w$ (mm) & 0.35 & 0.35 & 0.35 & 0.35 & 0.35 & 0.35 \\
      Thickness $t$ (mm) & 0.58 & 0.35 & 0.24 & 0.12 & 0.065 & 0.0085 \\
      $l/t$ & 11 & 19 & 20 & 38 & 37 & 329\\
      Surface area $S$ (mm$^2$) & 18.6 & 14.0 & 9.7 & 7.7 & 4.2 & 4.2\\\hline
    \end{tabular}
  \end{center}
\end{table}

\subsection*{Heat transport measurements}
The thermal conductivity was measured by using a home build system. We employed a standard one-heater-two-thermometers steady-state method.
A thermal gradient was applied along the sample by heating a chip resistor which was surrounded by a thin Ag foil. Heater power was determined by $IV$ where $I$ is applied current to the heater by using a dc current source (Yokogwa GS200) and $V$ is generated electric voltage which was measured by a digital multimeter (Keithley 2000). Two differential thermocouples were used to measure the temperature gradient. The thermocouple was made by spot welding the Chromel wire of 25 $\mu$m diameter directly to the Constantan wire of 25 $\mu$m diameter, and was glued on the sample by Dupont 4922N silver paste.
Thermoelectric voltage developed in thermocouples was measured by a digital nanovoltmeter (Keithley 2182A).
To minimize the effect of radiation, the setup for the thermal conductivity measurement was placed in a radiation shield tightly connected to the thermal bath.
The thermal conductivity was checked to be independent of the thermal gradient applied by changing $\Delta T/T$ in the range of 0.5–3 $\%$.
Main source of uncertainty in thermal conductivity is the loss of the applied heat by radiation through the heater and the sample as described below.
Another source of uncertainty results from
uncertainty in the measured thickness of the samples, which is about 6
$\%$ at maximum in the thinnest sample.

\section*{Supplementary Text}
\subsection*{Reliability of our thermal conductivity measurements}
In order to make sure that our experimental setup works properly, thermal
conductivity $\kappa$ of a thin Ag foil was measured. The sample has a bar shape with the length of 5 mm. The width and thickness are 1 and 0.1 mm, respectively. The result is shown by solid red circle in Figure~S1. 
$\kappa$ exhibits a typical textbook-behavior, namely nearly constant $\kappa$ at high temperatures, a profound conductivity maximum at low temperatures, and a steep decrease below the maximum following $\kappa\sim T$. In the inset of Figure~S1, the Lorenz ratio $L/L_0$ is plotted as a function of temperature where $L=(\kappa/T)\rho$ and $L_0$ is the Lorenz constant, $L_0=2.44\times10^{-8}$ W$\Omega$/K$^2$.
In metallic systems, at low and at high temperatures where elastic
scattering of electrons dominates, the Wiedemann-Franz (WF) law, stating that $L=L_0$, is satisfied in both limit.
As seen from the figure, the law is confirmed to be satisfied in Ag~\cite{jaoui} and Cu~\cite{zhang,berman} at low and at high temperatures, but is violated ($L<L_0$) in the intermediate temperature range due to
the presence of inelastic scattering by phonons.
Our result reasonably agrees with the literature data, indicating that the WF law is properly verified in the Ag foil by using our setup.

\subsection*{Quantifying heat loss by measuring a highly resistive metallic foil}

The heat loss along thermocouple wires and Manganin wires of 50 $\mu$m diameter connected to the heater are many orders of magnitude lower than the thermal current along the sample because thermal resistance of these wires~\cite{sundqvist,zavaritskii} are much larger than that of the graphite samples (Fig.~S2). The heat loss by residual gas is also negligible because the measurements were carried out in high vacuum. Heat leak  by radiation becomes important at high temperatures for samples with high thermal resistance. 

We quantified heat loss by testing the validity of the WF law in a second Ag foil with a higher thermal resistance, which was prepared by narrowing the width down to 0.3 mm, while keeping the length and thickness the same as the original 1 mm wide Ag foil. Hereafter, we call the Ag foils with the width of 1 and 0.3 mm as wide and narrow Ag foils, respectively.
In Figure~S1, we show thermal conductivity of the narrow Ag foil (open circles), which displays an upward deviation from the nearly constant $\kappa$ of the wide Ag foil above 100 K. Consequently, a small deviation from the WF law was resolved above 100 K, which is about 14 $\%$ at 300 K (the inset of Fig.~S1). This deviation is attributed to the radiation loss of the applied heater power. Note that the thermal resistance of the narrow Ag foil is comparable with the one of the 8.5 $\mu$m graphite sample as shown in Figure~S3.

From the difference in $\kappa$ between the narrow and wide Ag foils, we estimated thermal conductance $K_{\rm loss}$ of radiation as follows;
\begin{equation}
    K_{\rm loss}=(\kappa_{\rm narrow}-\kappa_{\rm wide})\cdot G,
\end{equation}
where $G$ is a geometric factor of the narrow Ag foil.
Temperature dependence of $K_{\rm loss}$ is shown in Figure~S4A.

\subsection*{Estimation of the radiative heat loss}

The radiative thermal conductance $K_{\rm rad}$ from the hot part of the setup to the surrounding isothermal shield is given by:
\begin{equation}
K_{\rm rad}=\frac{\sigma_{\rm T}\varepsilon S(T_{\rm hot}^4-T_{\rm shield}^4)}{\Delta T},
\end{equation}
where $\sigma_{\rm T}$ is Stefan-Boltzmann constant, $T_{\rm hot}$ is the temperature of the hot part of the setup, $\varepsilon$ and $S$ are the emissivity and the surface area, respectively. $T_{\rm shield}$ is the temperature of the radiation shield and $\Delta T$ is the temperature difference determined by the thermocouples. Heat can be radiated from both the heater and the Ag sample, so $T_{\rm hot}$ can be either $T_{\rm heater}$ or $T_{\rm Ag}$. Let us discuss which can be the dominant source of heat loss.\\
In our setup, the heater is surrounded by the thin Ag foil with surface area of $S_{\rm heater}$ = 150 mm$^2$. This foil and the Ag samples used for the thermal conductivity measurements were cut from the same Ag sheet. A crude approximation of its emissivity is $\varepsilon_{\rm Ag}$ = 0.05 which is a typical value for metallic surfaces. The surface area of the wide (narrow) Ag foil sample is $S_{\rm Ag}$ = 11 (4) mm$^2$. The fact that the surface area of the heater is more than one order of magnitude larger than the Ag sample ($S_{\rm heater}\gg S_{\rm Ag}$) and the  heater is hotter than the Ag sample ($T_{\rm heater}>T_{\rm Ag}$) imply that the radiative heat conduction from the heater dominates over the one from the Ag sample ($K_{\rm rad}^{\rm heater}\gg$ $K_{\rm rad}^{\rm Ag}$).\\
In this case, when measuring the Ag foils, the heat loss expressed in Eq.~(1) is equal to the radiation loss from the heater and therefore:
\begin{equation}
    K_{\rm loss}\sim K_{\rm rad}^{\rm heater}=\frac{\sigma_{\rm T}\varepsilon_{\rm Ag} S_{\rm heater}(T_{\rm heater}^4-T_{\rm shield}^4)}{\Delta T}
\end{equation}
Let us now discuss the heat loss in the case of graphite. Here important points are i) the surface areas of the graphite samples are also much smaller than the heater (see Table~S1), ii) the thermal resistance of the most resistive (thinnest) graphite sample is comparable with the narrow Ag foils, and iii) the emissivity of graphite is higher than Ag. In order to allow the highest experimental margin, let us assume the maximum emissivity of $\varepsilon_{\rm graphite}$ = 1. By taking into account these points, we estimate the heat loss by the two ways. \\
One is that the heat is lost from the heater as in the case of the Ag foils based on the facts i) and ii). In this case,  an identical $K_{\rm loss}$ estimated from Eq.~(1) is subtracted from the measured thermal conductance $K_{\rm measure}$ of all graphite samples and the thermal conductivity would be:
\begin{equation}
    \kappa=(K_{\rm measured}-K_{\rm loss})\cdot\frac{l}{wt},
\end{equation}
where the length $l$, width $w$, and thickness $t$ of each graphite samples are shown in Table~S1.
This estimation gives upper limit of error in the thermal conductivity shown in Figure~3.
However, heat can also be lost from the sample surfaces because of the higher emissivity of graphite. In this case, $\kappa$ should be  corrected as :
\begin{equation}
    \kappa=(K_{\rm measure}-K_{\rm loss}-K_{\rm rad}^{\rm graphite})\cdot\frac{l}{wt},
\end{equation}
where the radiative heat conduction from the graphite to the shield $K_{\rm rad}^{\rm graphite}$ is evaluated as,
\begin{equation}
    K_{\rm rad}^{\rm graphite}=K_{\rm rad}^{\rm heater}\cdot\frac{\varepsilon_{\rm graphite}}{\varepsilon_{\rm Ag}}\cdot\frac{S_{\rm graphite}}{S_{\rm heater}}\sim K_{\rm loss} \cdot\frac{\varepsilon_{\rm graphite}}{\varepsilon_{\rm Ag}}\cdot\frac{S_{\rm graphite}}{S_{\rm heater}}.
\end{equation}
Figure 4B shows $K_{\rm rad}^{\rm graphite}$ for different graphite samples. Obviously, $K_{\rm rad}^{\rm graphite}$ overestimate the radiative heat loss from the graphite sample, because the sample is colder than the heater ($T_{\rm heater}>T_{\rm graphite}$). Thus, this estimation gives an upper boundary to the thermal conductivity shown in Figure~3.

Figure~S5 show (A) the total thermal conductance $K$, (B) the radiative thermal conductance from the heater and the the graphite sample $K_{\rm loss}+K_{\rm rad}^{\rm graphite}$, and (C) fraction of $K_{\rm loss}+K_{\rm rad}^{\rm graphite}$ in the total thermal conductance at 250 K for the various thickness graphite samples. 
One can see from the figures that relative heat loss decreases with decreasing thickness reflecting the reduction in surface areas, pointing that the radiation loss is less effective in thinner samples. The only exception is the 8.5 $\mu$m sample. However, even in this case, the fraction of heat loss remains less than 25 $\%$. Thus the observed evolution of the thermal conductivity with decreasing thickness cannot be attributed to a change in the heat loss by radiation.

\subsection*{Specific heat}
Specific heat $C$ of graphite was measured between 2 K and 300 K using a relaxation method of the heat capacity option in a Quantum Design, PPMS instrument. 
A single 3.6-mg piece of graphite was used, which was cut from the same mother crystal used for the heat transport experiments.
Our specific heat data agree well with the previous report~\cite{desorbo,alexander} as shown in Figure~S6. 
Below 10 K, the specific heat follows a $T^{2.5}$ behavior as previously reported~\cite{alexander}.
Our result also coincides with the calculated specific heat by Nihira~\cite{nihira}
The specific heat data was used to evaluate the thermal diffusivity shown in Figures~2(C) and 3(B).

\subsection*{Comparison to the literature data}
In Figure~S7(A), thermal conductivity of the 580 $\mu$m thickness sample is shown together with the previously reported data by Bowman~\cite{bowman} and Holland~\cite{holland}.
Three data are roughly coincide with each other.
We estimated the thermal diffusivity from the literature data of $\kappa$~\cite{bowman,holland} by using the specific heat data by DeSorbo~\cite{desorbo} and Alexander~\cite{alexander}, which are shown in Figure~S7(B). 
It is remarkable that the Poiseuille maximum and Knudsen minimum can be can found in published data, but they have remained unnoticed. 

\subsection*{Electric resistivity}
Electric resistivity measurement was done by a standard four-point technique
using the same contacts and wires utilized for the thermal conductivity measurements.
The result for the 580 $\mu$m thickness sample is shown in Figure~S8, which was used for the evaluation of Lorenz ratio $L/L_0$ shown in Figure~2(D).

\clearpage
\begin{figure}
\vspace*{-7cm}
\begin{center}
\includegraphics[width=15cm]{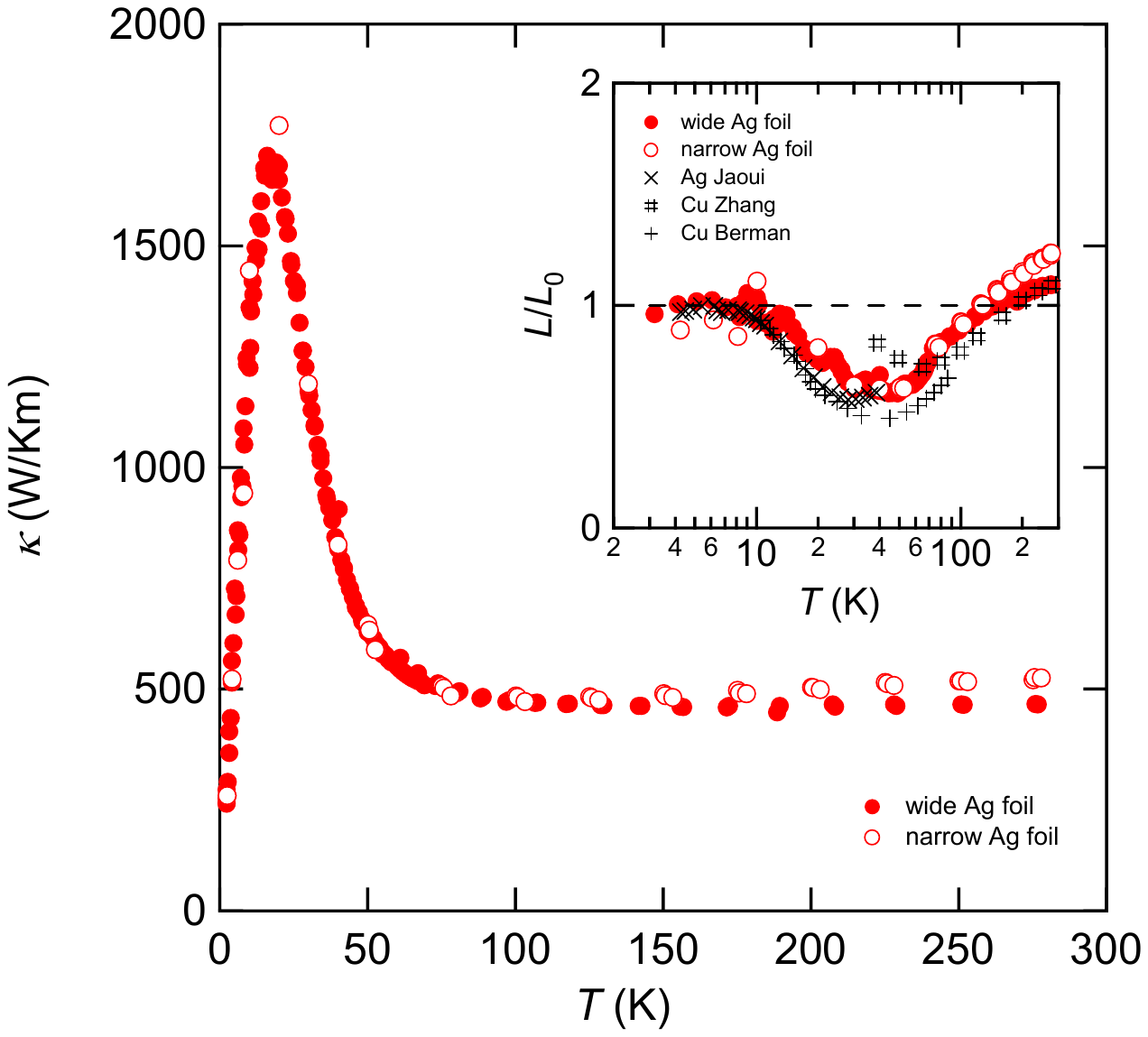}\\
\end{center}
\vspace*{-6cm}
{\bf Fig. S1: Thermal conductivity of silver.} Temperature dependence of thermal conductivity $\kappa$ of two different Ag foils with different width. The data for the wide and narrow Ag foils are represented by closed and open circles, respectively. The Lorenz ratios $L/L_0$ are shown in the inset. $L/L_0$ for the wide Ag foil reasonably agrees with the literature data of Ag~\cite{jaoui} and Cu~\cite{zhang,berman}, but a small deviation from the Wiedemann-Franz law is resolved in the narrow Ag foil above 100 K.
\end{figure}

\clearpage

\begin{figure}
\vspace*{-7cm}
\begin{center}
\includegraphics[width=15cm]{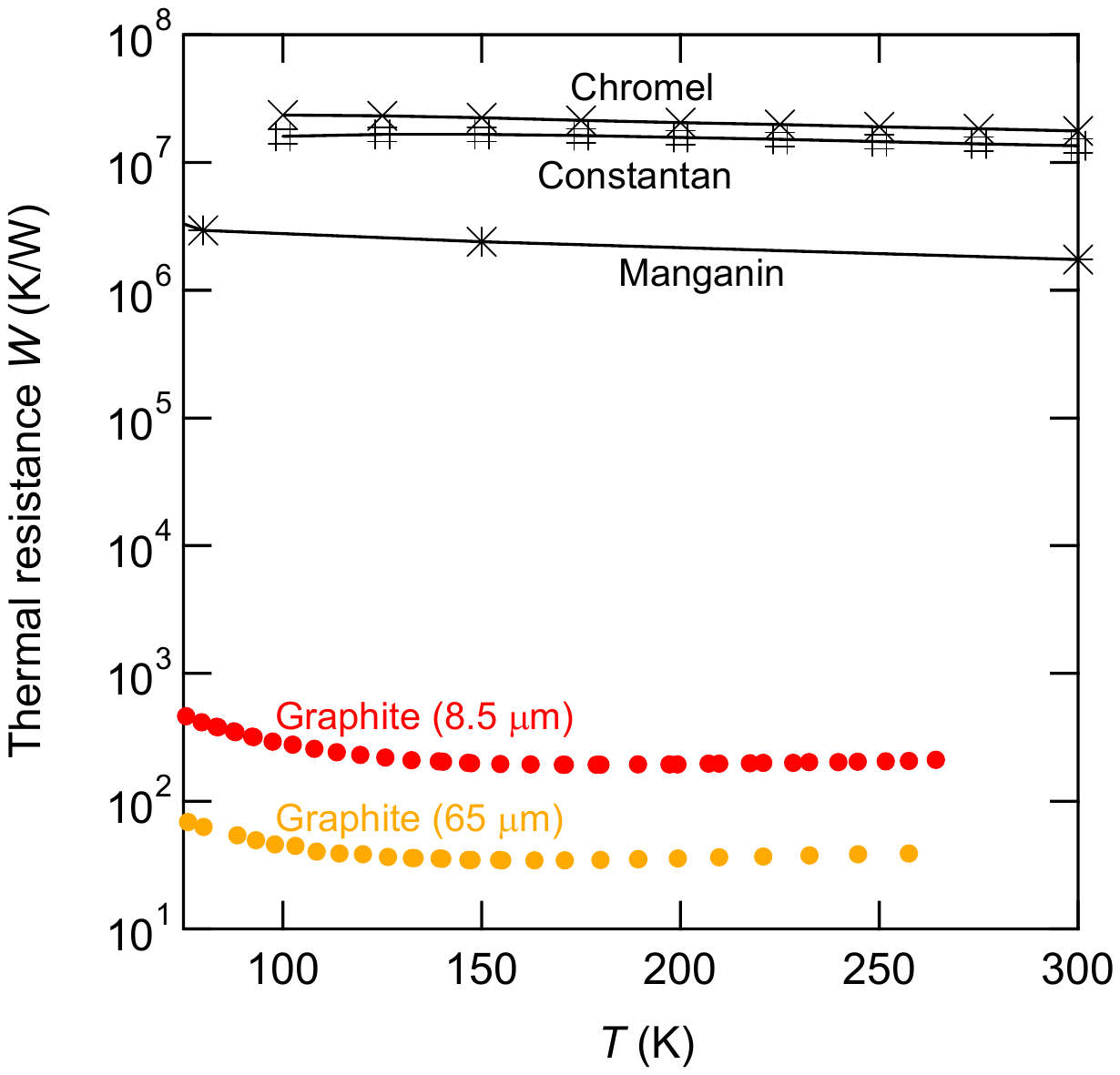}\\
\end{center}
\vspace*{-6cm}
{\bf Fig. S2: Thermal resistance of graphite samples and wires.} Temperature dependence of thermal resistance of the graphite samples with the thickness of 8.5 and 65 $\mu$m and wires used in our setup; Manganin, Constantan, and Chromel wires~\cite{sundqvist,zavaritskii}.
\end{figure}

\clearpage

\begin{figure}
\vspace*{-7cm}
\begin{center}
\includegraphics[width=15cm]{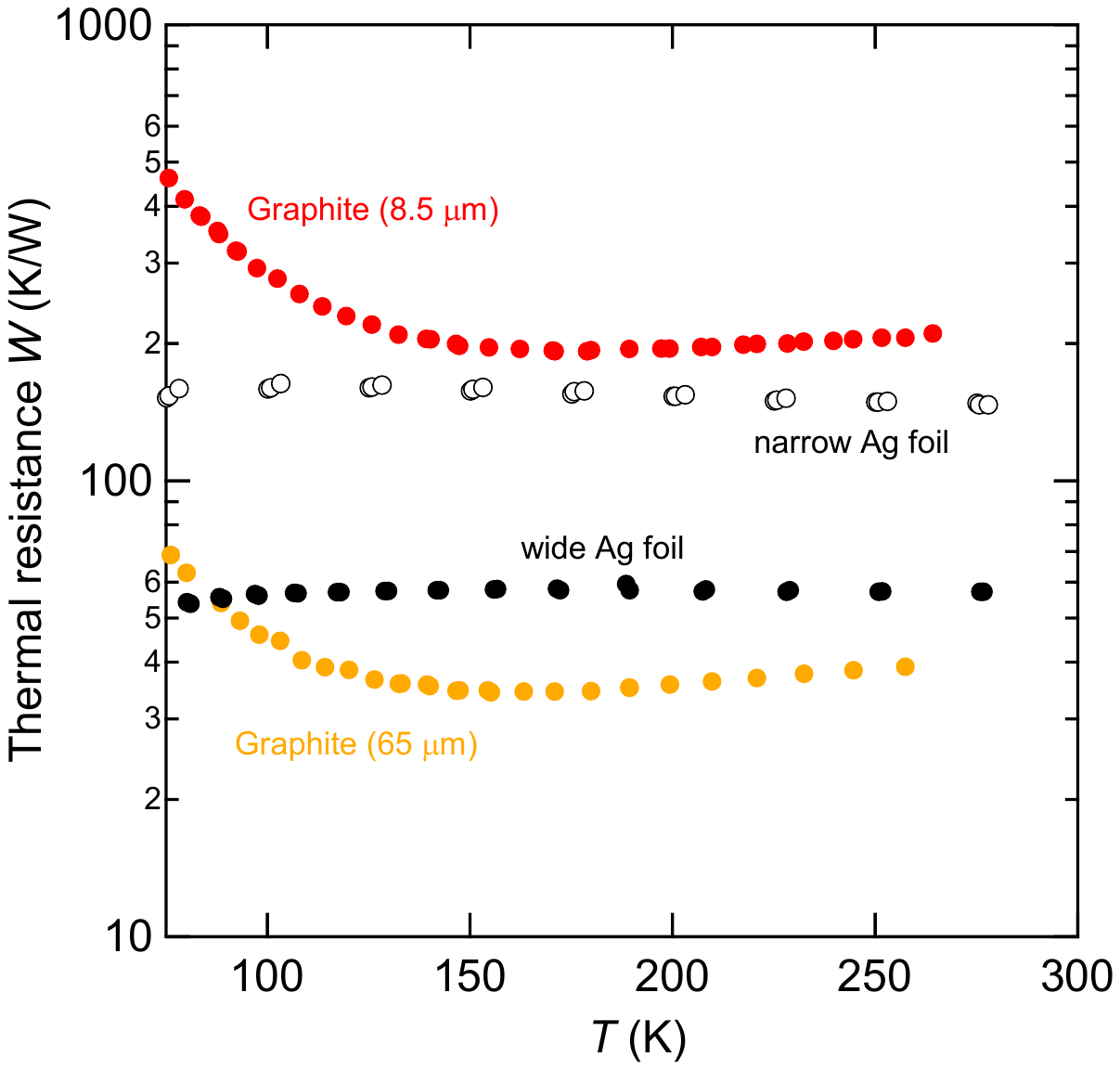}\\
\end{center}
\vspace*{-6cm}
{\bf Fig. S3: Thermal resistance of graphite samples and Ag foils.} Temperature dependence of thermal resistance $W$ of the graphite samples with the thickness of 8.5 and 65 $\mu$m and the narrow and wide Ag foils. $W$ of the narrow Ag foil is comparable with the one of the 8.5 $\mu$m graphite sample.
\end{figure}

\clearpage

\begin{figure}
\vspace*{-4cm}
\begin{center}
\includegraphics[width=15cm]{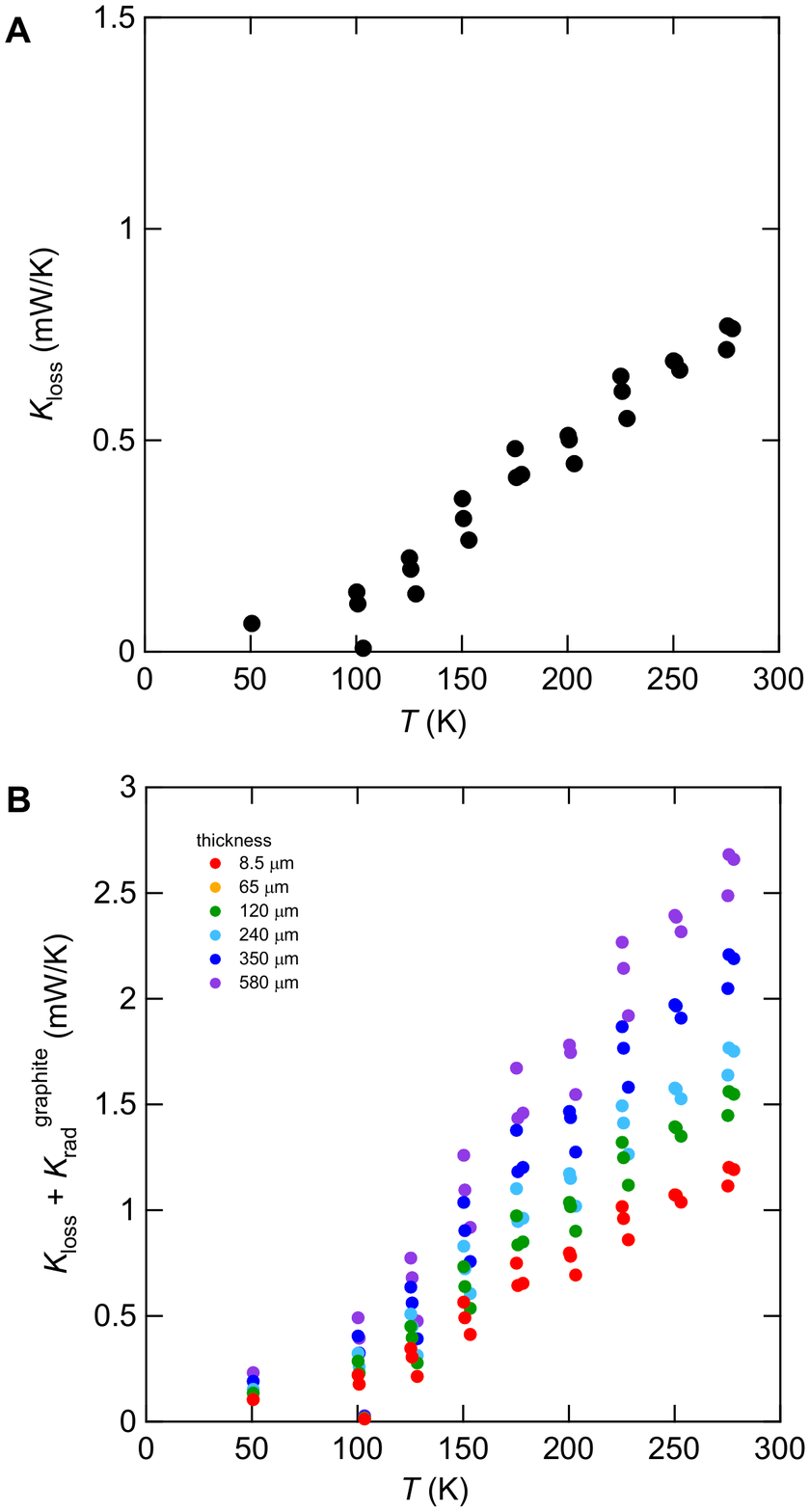}\\
\end{center}
\vspace*{-2cm}
{\bf Fig. S4: Thermal conductance by heat loss.} (A) Heat loss thermal conductance measured by comparing the deviation from the Wiedemann-Franz law in two Ag foils.(A) This is the minimum correction to the measured thermal conductance assuming that the radiation loss is due to the heater and therefore identical for different samples. (B) Heat loss thermal conductance for different graphite samples assuming important radiation from the sample itself in addition to the radiation loss from the heater. The loss is larger in samples with larger surfaces.
\end{figure}

\clearpage

\begin{figure}
\vspace*{-2cm}
\begin{center}
\includegraphics[width=15cm]{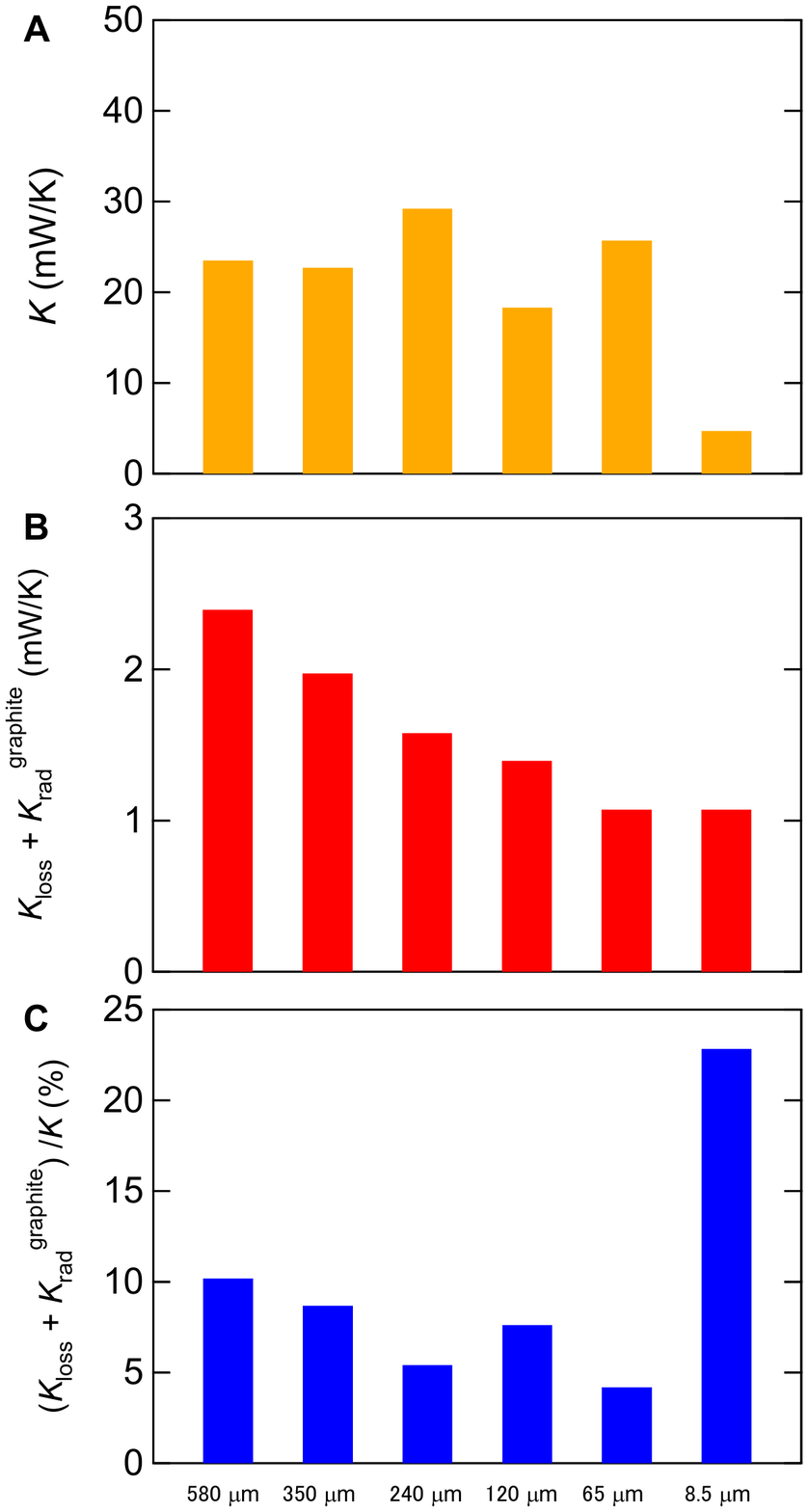}\\
\end{center}
\vspace*{-2cm}
{\bf Fig. S5: Thermal conductance of graphite samples at 250 K and the relative weight heat loss.} (A) Thermal conductance $K$ of graphite samples and (B) radiative thermal conductance from the heater and the graphite sample $K_{\rm loss}+K_{\rm rad}^{\rm graphite}$ with different thicknesses. (C) Relative correction to the thermal conductance due to radiative heat loss taking in to account the surface of each graphite sample as well as the heater.
\end{figure}

\clearpage

\begin{figure}
\vspace*{-7cm}
\begin{center}
\includegraphics[width=15cm]{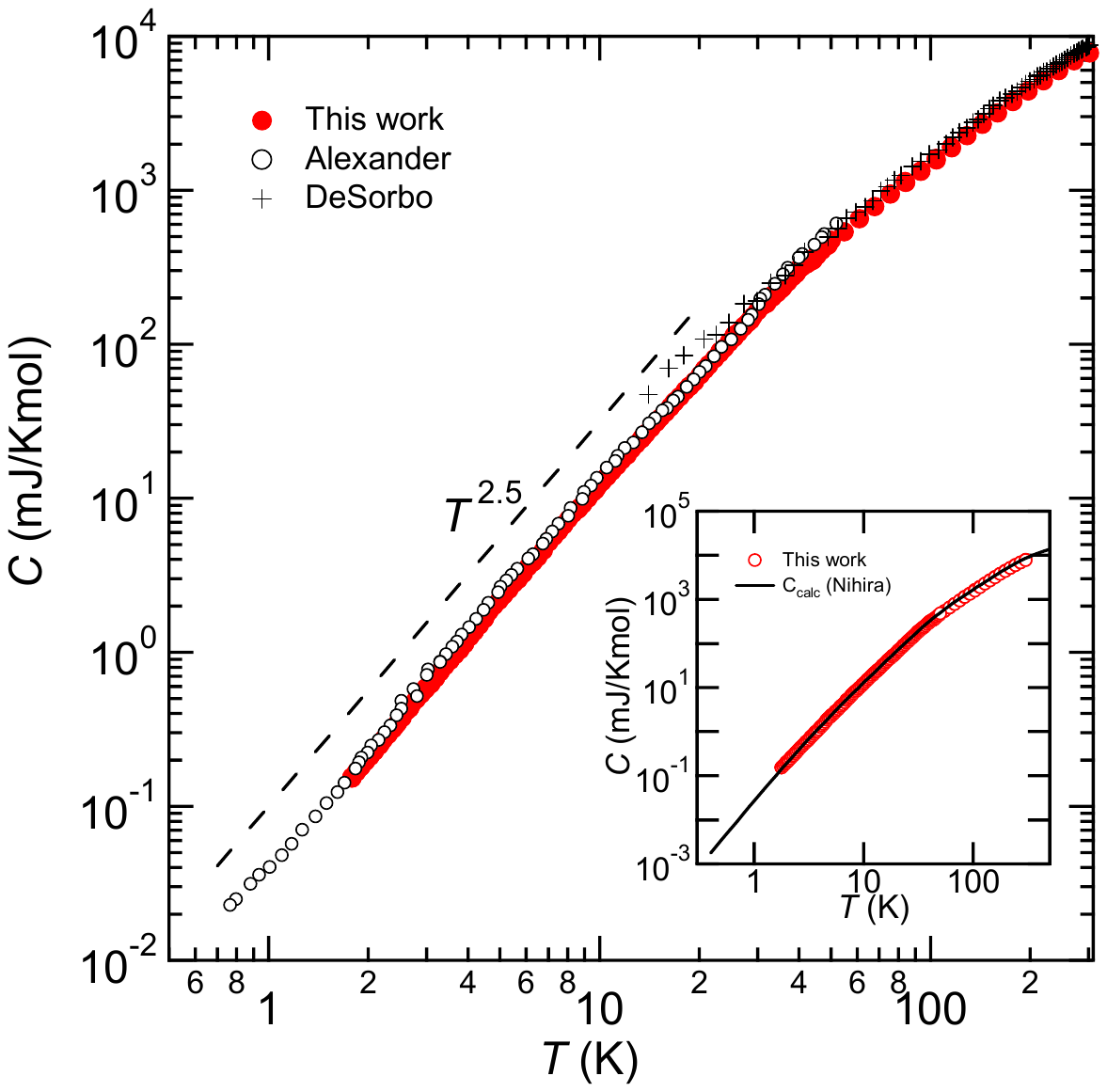}\\
\end{center}
\vspace*{-6cm}
{\bf Fig. S6: Specific heat.} Temperature dependence of specific heat $C$ of graphite in logarithmic scale with the literature data~\cite{desorbo,alexander}. Our specific heat data is compared with the calculated $C$ by Nihira~\cite{nihira}.
\end{figure}

\clearpage

\begin{figure}
\vspace*{-5cm}
\begin{center}
\includegraphics[width=13cm]{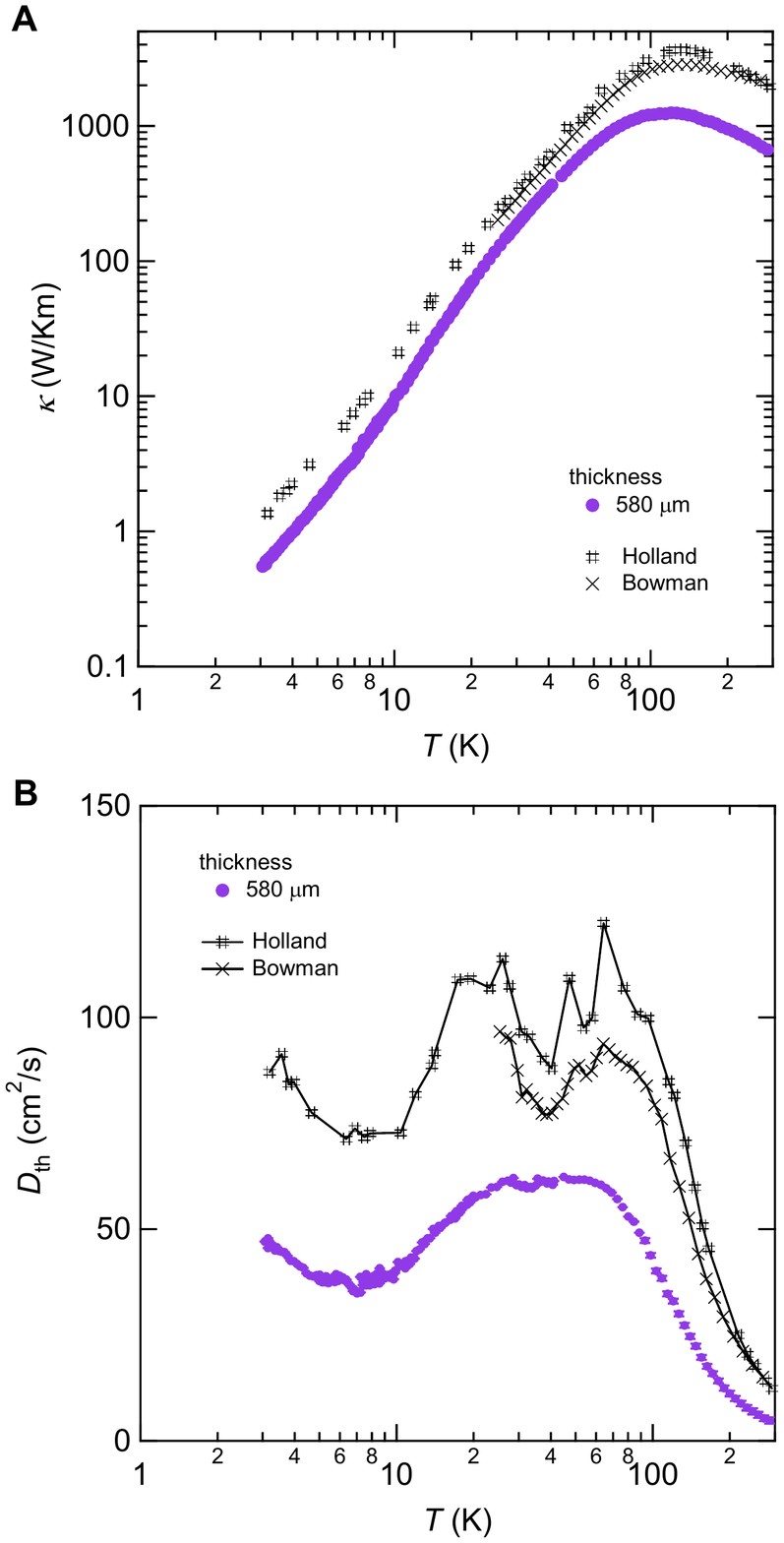}\\
\end{center}
\vspace*{-1cm}
{\bf Fig. S7: Comparison to the literature data.} Our thermal conductivity (A) and thermal diffusivity (B) data for the 580 $\mu$m sample are compared with the literature data~\cite{desorbo,alexander}.
\end{figure}

\clearpage

\begin{figure}
\vspace*{-7cm}
\begin{center}
\includegraphics[width=15cm]{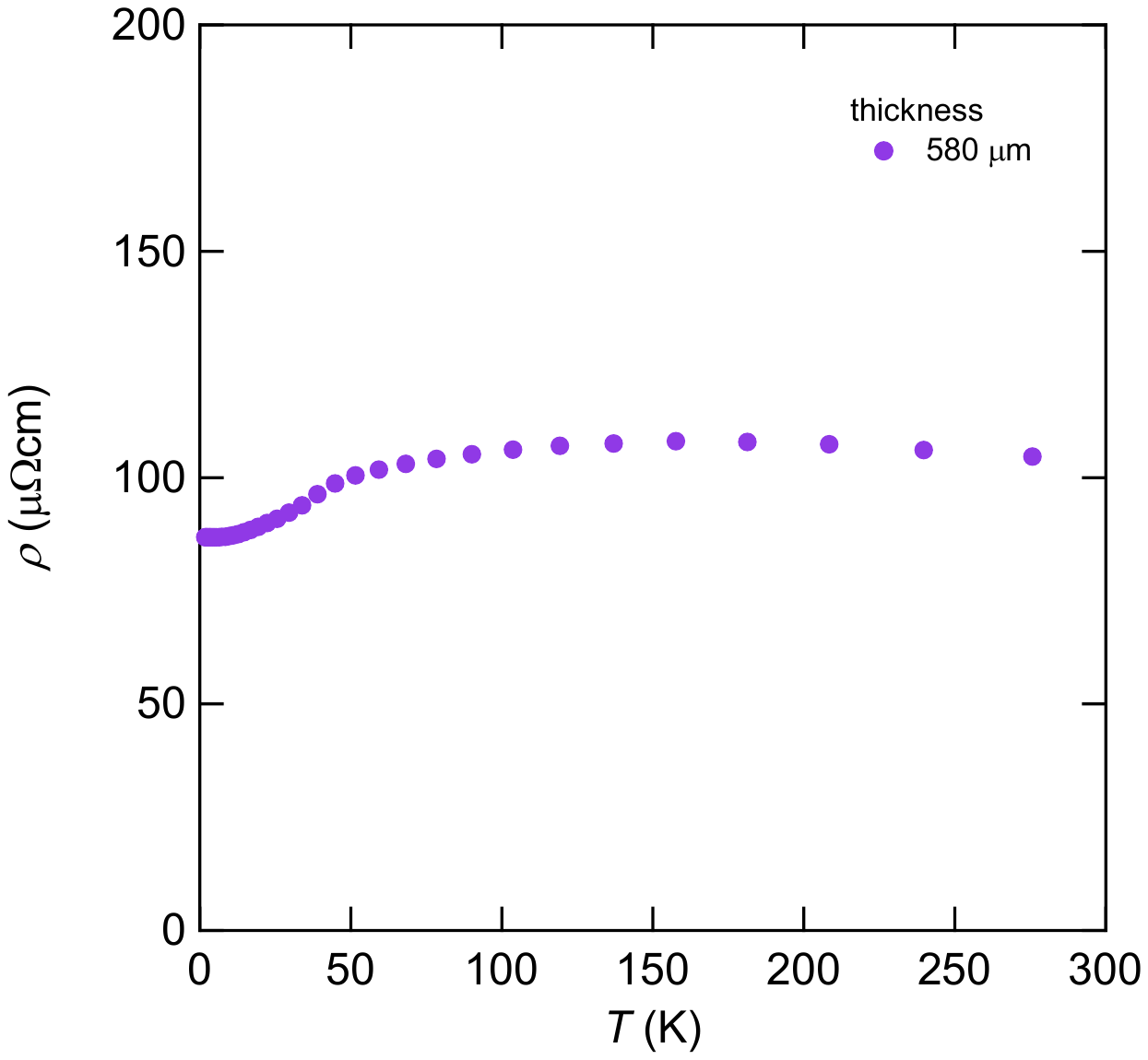}\\
\end{center}
\vspace*{-6cm}
{\bf Fig. S8: Electric resistivity.} Temperature dependence of electric resistivity $\rho$ of the 580 $\mu$m thickness graphite sample.
\end{figure}

\begin{thebibliography}{99}

\bibitem{ziman} J. M. Ziman, \textit{Electrons and Phonons: The Theory of Transport Phenomena in Solids} (Oxford Univ. Press, 2001). 
\bibitem{gurzhi} R. N. Gurzhi, Hydrodynamic effects in solids at low temperature, Sov. Phys.-Usp. \textbf{11}, 255-270 (1968).
\bibitem{bandurin} D. A. Bandurin, I. Torre, R. K. Kumar, M. B. Shalom, A. Tomadin, A. Principi, G. H. Auton, E. Khestanova, K. S. Novoselov, I. V. Grigorieva, L. A. Ponomarenko, A. K. Geim, M. Polini, Negative local resistance caused by viscous electron backflow in graphene, Science \textbf{351}, 1055-1058 (2016). 
\bibitem{crossno} J. Crossno, Jing K. Shi, K. Wang, X. Liu, A. Harzheim, A. Lucas, S. Sachdev, P. Kim, T. Taniguchi, K. Watanabe, T. A. Ohki, K. C. Fong, Observation of the Dirac fluid and the breakdown of the Wiedemann-Franz law in graphene, Science \textbf{351}, 1058-1061 (2016). 
\bibitem{moll} P. J. W. Moll, P.  Kushwaha, N. Nandi, B. Schmidt, A. P. Mackenzie, Evidence for hydrodynamic electron flow in PdCoO$_2$, Science \textbf{351}, 1061-1064 (2016).
\bibitem{lee} S. Lee, D. Broido, K. Esfarjani, G. Chen, Hydrodynamic phonon transport in suspended graphene. Nat. Commun. \textbf{6}, 6290 (2015).
\bibitem{cepellotti} A. Cepellotti, G. Fugallo, L. Paulatto, M. Lazzeri, F. Mauri, N. Marzaria, Phonon hydrodynamics in two-dimensional materials. Nat. Commun. \textbf{6}, 6400 (2015).
\bibitem{machida} Y. Machida, A. Subedi, K. Akiba, A. Miyake, M. Tokunaga, Y. Akahama, K. Izawa, K. Behnia, Observation of Poiseuille flow of phonons in black phosphorus, Sci. Adv. \textbf{4}, eaat3374 (2018).
\bibitem{martelli} V. Martelli, J. L. Jimnez, M. Continentino, E. Baggio-
Saitovitch, K. Behnia, Thermal transport and phonon hydrodynamics in strontium titanate, Phys. Rev. Lett. \textbf{129}, 125901 (2018).
\bibitem{huberman} S. Huberman, R. A. Duncan, K. Chen, B. Song, V. Chiloyan, Z. Ding, A. A. Maznev, G. Chen, K. A. Nelson, Observation of second sound in graphite at temperatures above 100 K, Science \textbf{364}, 375-379 (2019).
\bibitem{deglin} L. P. Mezhov-Deglin, Measurement of the thermal conductivity of crystalline {H}e$^4$, Zh. Eksp. Teor. Fiz. \textbf{49}, 66-79 (1965).
\bibitem{thomlinson} W. C. Thomlinson, Evidence for anomalous phonon excitations in solid {H}e$^3$, Phys. Rev. Lett. \textbf{23}, 1330-1332 (1969).
\bibitem{kopylov} V. N. Kopylov and L. P. Mezhov-Deglin, Investigation of the kinetic coefficients of bismuth at helium temperatures, Zh. Eksp. Teor. Fiz. \textbf{65}, 720-734 (1973).
\bibitem{zholonko} N. N. Zholonko, Poiseuille flow of phonons in solid hydrogen, Phys. Solid State \textbf{48}, 1678-1680 (2006).
\bibitem{nika} D. L. Nika, A. A. Balandin, Phonons and thermal transport in graphene and graphene-based materials, Rep. Prog. Phys. \textbf{80}, 036502 (2017). 
\bibitem{ding} Z. Ding, J. Zhou, B. Song, V. Chiloyan, M. Li, T-H. Liu, G. Chen, Phonon hydrodynamic heat conduction and Knudsen minimum in
graphite, Nano Lett. \textbf{18}, 638-649 (2018).
\bibitem{geim}
A. K. Geim  and K. S. Novoselov, The rise of graphene, Nat. Mater. \textbf{6}, 183-191 (2007).
\bibitem{krumhansl} J. A. Krumhansl and H. Brooks, The lattice vibration specific heat of graphite, J. Chem. Phys. \textbf{21}, 1663-1669 (1953).
\bibitem{slack} G. N. Slack, Anisotropic thermal conductivity of pyrolytic graphite, Phys. Rev. \textbf{127}, 694-701 (1962).
\bibitem{bowman} J. C. Bowman, J. A. Krumhansl, J. T. Meers, \textit{Industrial Carbon and Graphite} (Society of Chemical Industry, London, 1958), p. 52.
\bibitem{holland} M. G. Holland, C. A. Klein, W. D. Straub, The Lorenz number of graphite at very low temperatures, J. Phys. Chem. Solids \textbf{27}, 903-906 (1966).
\bibitem{taylor} R. Taylor, The thermal conductivity of pyrolytic graphite, Phil. Mag. \textbf{13}, 157-166 (1966).
\bibitem{morelli} D. T. Morelli, C. Uher, Thermal conductivity and thermopower of graphite at very low temperatures, Phys. Rev. B \textbf{31}, 67216725 (1985).
\bibitem{SM} Materials and Methods are available as Supplementary Materials on Science Online.
\bibitem{alexander} M. G. Alexander, D. P. Goshorn, D. G. Onn, Low-temperature specific heat of the graphite intercalation compounds KC$_8$, CsC$_8$, RbC$_8$ and
their parent highly oriented pyrolytic graphite, Phys. Rev. B \textbf{22}, 4535-4542 (1980).
\bibitem{komatsu} K. Komatsu, Theory of the specific heat of graphite II, J. Phys. Soc. Jpn. \textbf{10}, 346-356 (1955).
\bibitem{chen} S. Chen, Q. Wu, C. Mishra, J. Kang, H. Zhang, K. Cho, W. Cai, A. A. Balandin, R. S. Ruoff, Thermal conductivity of isotopically modified graphene, Nat. Mater. \textbf{11}, 203-207 (2012).
\bibitem{wei} L. Wei, P. K. Kuo, R. L. Thomas, T. R. Anthony, W. F. Banholzer, Thermal conductivity of isotopically modified single crystal diamond, Phys. Rev. Lett. \textbf{70}, 3764-3767 (1993).
\bibitem{kang} J. S. Kang, M. Li, H. Wu, H. Nguyen, Y. Hu,  Experimental observation of high thermal conductivity boron arsenide, Science \textbf{361}, 575-578 (2018).
\bibitem{li} S. Li, Q. Zheng, Y. Lv, X. Liu, X. Wang,
P. Y. Huang, D. G. Cahill, B. Lv, High thermal conductivity in cubic boron arsenide crystals, Science \textbf{361}, 579-581 (2018).
\bibitem{tian} F. Tian, B. Song, X. Chen, N. K. Ravichandran, Yinchuan Lv, K. Chen, S. Sullivan, J. Kim, Y. Zhou, T.-H. Liu, M. Goni,
Z. Ding, J. Sun, G. A. G. U. Gamage, H. Sun,
H. Ziyaee, S. Huyan, L. Deng, J. Zhou, A. J. Schmidt,
S. Chen, C.-W. Chu, P. Y. Huang, D. Broido, Li Shi,
G. Chen, Z. Ren, Unusual high thermal conductivity in boron arsenide bulk crystals, Science \textbf{361}, 582-582 (2018).
\bibitem{balandin} A. A. Balandin, S. Ghosh, W. Bao, I. Calizo, D. Teweldebrhan, F. Miao, C. N. Lau, Superior thermal conductivity of single-layer graphene, Nano Lett. \textbf{8}, 902-907 (2008).
\bibitem{nihira} T. Nihira, T. Iwata, Temperature dependence of lattice vibrations and analysis of the specific heat of graphite, Phys. Rev. B \textbf{68}, 134305 (2003).
\bibitem{nicklow} R. Nicklow and N. Wakabayashi, H. G. Smith, Lattice dynamics of phyrolytic graphite, Phys. Rev. B \textbf{5}, 4951-4962 (1972).
\bibitem{cong} X. Cong, Q.-Q. Li, X. Zhang, M.-L. Lin, J.-B. Wu, X.-L. Liu,
P. Venezuel, P.-H. Tan, Probing the acoustic phonon dispersion and sound velocity of
graphene by Raman spectroscopy, Carbon \textbf{149}, 19-24 (2019).
\bibitem{jaoui} A. Jaoui, B. Fauqu\'{\rm e}, C. W. Rischau, A. Subedi, C. Fu, J. Gooth, N. Kumar, V. Süß, D. L. Maslov, C. Felser, K. Behnia, Departure from the Wiedemann–Franz law in WP$_2$ driven by mismatch in $T$-square resistivity prefactors, npj Quantum Materials \textbf{3}, 64 (2018).
\bibitem{zhang} Y. Zhang, N. P. Ong, Z. A. Xu, K. Krishana, R. Gagnon, L. Taillefer, Determining the Wiedemann-Franz ratio from the thermal Hall conductivity: Application to Cu and YBa$_2$Cu$_3$O$_{6.95}$, Phys. Rev. Lett. \textbf{84}, 2219-2222 (2000).
\bibitem{berman} R. Berman, D. K. C. Macdonald, The thermal and electrical conductivity of copper at low temperatures, Proc. R. Soc. London A \textbf{211}, 122-128 (1952).
\bibitem{sundqvist} B. Sundqvist, Thermal diffusivity and thermal conductivity of Chromel, Alumel, and Constantan
in the range 100–450 K, J. Appl. Phys. \textbf{72}, 539-545 (1992).
\bibitem{zavaritskii} N. V. Zavaritskii, A. G. Zeldovich,  Thermal conductivity of some technical materials at low temperatures, Zhur. Tekh. Fiz. \textbf{26}, 2032, (1956).
\bibitem{desorbo} W. DeSorbo, W. W. Tyler, The specific heat of graphite from 13 to 300 K, J. Chem. Phys. \textbf{21}, 1660-1663 (1953).
\end{thebibliography}
\end{document}